\documentstyle[12pt]{article}

\begin{document}

\begin{flushright}
IMSc/2005/09/22 
\end{flushright} 

\vspace{2mm}

\vspace{2ex}

\begin{center}
{\large \bf Branes in a Time Dependent Universe } \\ 



\vspace{8ex}

{\large  S. Kalyana Rama}

\vspace{3ex}

Institute of Mathematical Sciences, C. I. T. Campus, 

Taramani, CHENNAI 600 113, India. 

\vspace{1ex}

email: krama@imsc.res.in \\ 

\end{center}

\vspace{6ex}

\centerline{ABSTRACT}
\begin{quote} 

Long ago, McVittie had found a class of solutions which can be
thought of as Schwarzschild black holes in an FRW universe. In
recent years they have been studied extensively and generalised
to charged and uncharged black holes in $D \ge 4$ dimensions
also. Here, assuming an ansatz similar to McVittie's, we present
solutions for uncharged branes which can be thought of as branes
in a time dependent universe. We consider their application to
the brane antibrane decay process, also referred to as tachyon
condensation, and discuss the necessary generalisations required
for our ansatz to describe such a process.

\end{quote}

\vspace{2ex}


\newpage

{\bf 1.}  
Static brane solutions in string/M theory are well known
\cite{solns}. It is of obvious interest to study time dependent
solutions also. In recent years there have been numerous studies
of strings and branes in a variety of time dependent backgrounds
\cite{timesolns}.

Here we consider a different class of solutions, of the type
McVittie found long ago for four dimensional Schwarzschild black
holes \cite{mcvittie}. These McVittie solutions are
inhomogeneous and time dependent, and are thought of as
describing black holes in an FRW universe with perfect fluid
matter. These solutions and their properties have been discussed
extensively in \cite{nolan, book, sing} and are also generalised
to both charged and uncharged black holes in $D \ge 4$
dimensions \cite{sv, dad, gao}.

In this paper, we study brane solutions in a time dependent
universe by generalising the McVittie ansatz suitably. We
consider only the uncharged branes since the appropriate
generalisation to the charged case is not clear to us at
present. However we set up the full set of equations of motion,
corresponding to a sufficiently general metric applicable to the
charged case also. We then present the solutions for the
uncharged case which reduce to the known ones in suitable limits
and which may be thought of as decribing branes in a time
dependent universe. They may also be thought of as describing
branes with a time dependent mass. We illustrate this
possibility in a simplified context by making liberal
assumptions.

In string/M theory context, the static brane solutions are often
interpreted as coincident stacks of branes and antibranes
\cite{d}. In this context, see also \cite{multi} where more
general static multiparameter solutions and their interpretation
in terms of tachyon condensation are studied. The brane
antibrane system has a tachyon mode indicating the instability
due to brane antibrane pair annihilation. These pairs are
expected to annihilate totally \cite{ashoke}. This time
dependent decay process, also referred to as tachyon
condensation, must have a gravity description atleast in the
initial stages when the number of branes and antibranes is
sufficiently large. It is then natural to hope that the present
ansatz, suitably generalised, may describe such a process. We
present a discussion of the necessary generalisations which,
hopefully, are also sufficient.

The organisation of this paper is as follows. In section {\bf 2}
we present the equations of motion taking a sufficiently general
metric. In section {\bf 3}, we specialise to an ansatz similar
to McVittie's, present the solutions and discuss various limits
where they reduce to the known ones. In section {\bf 4} we
discuss these solutions as describing branes with a time
dependent mass. In section {\bf 5} we conclude with a discussion
of the necessary generalisations of the present ansatz required
to provide a gravity description of the brane decay process.

\vspace{4ex}

{\bf 2.}  
Consider the following $D-$dimensional action for the metric
$G_{M N}$, the scalar field $\phi$, a $(P + 2)-$form gauge field
strength $F$, and matter fields:
\begin{equation}\label{action} 
S = \frac{1}{2 \kappa^2} \; \int d^D x \sqrt{- G} 
\left( R - \frac{1}{2} \; \partial_M \phi \; \partial^M \phi 
- \frac{e^{\lambda \phi} \; F^2}{2 (P + 2) !} \; \right) \; 
+ \; S_{mat}
\end{equation} 
where $G = det \; G_{M N}$, $\lambda$ is a constant,
\[
F^2 = G^{M N} \; F^2_{M N} \; \; , \; \; \; \; 
F^2_{M N} = F_{M M_2 \cdots M_{P + 2}} \; 
F_N^{\; \; M_2 \cdots M_{P + 2}} \; , 
\] 
and $S_{mat}$ is the action for matter fields which is assumed
to be independent of $\phi$ and $F$ and has the energy momentum
tensor $T_{M N}$ given by
\[ 
\delta S_{mat} = - \frac{1}{2} \int d^D x \sqrt{- G} \; 
T_{M N} \; \delta G^{M N} \; .
\] 
The $D-$dimensional Newton's constant $G_D$ is given by
$\kappa^2 = 8 \pi G_D$. For $D = 11$, $\lambda = 0$, and $\phi$
absent, the above action describes M theory branes, whereas for
$D = 10$ and $\lambda = - 1$, $+ 1$, $\frac{3 - P}{2}$ it
describes fundamental strings, $NS \; 5-$branes, and $D
P-$branes respectively. The equations of motion that follow from
(\ref{action}) can be written as
\begin{eqnarray} 
& & R_{M N} = \frac{1}{2} \partial_M \phi \partial_N \phi 
+ \; S^F_{M N} + \; \kappa^2 S_{M N} \label{G} \\
& & \nabla^2 \phi = \frac{\lambda e^{\lambda \phi}} 
{2 (P + 2) !} \; F^2 \label{phi} \\
& & \partial_M \left( \sqrt{- G} \; e^{\lambda \phi} \; 
F^{M M_2 \cdots M_{P + 2}} \right) = 0 \label{F} \\
& & \nabla_M T^{M N} = 0 \label{t=0} 
\end{eqnarray} 
where the source terms $S^F_{M N}$ and $S_{M N}$ are given by 
\begin{eqnarray} 
& & S^F_{M N} = \frac{e^{\lambda \phi}}{2 (P + 2) !}  
\left( (P + 2) \; F^2_{M N} - \frac{P + 1}{D - 2}
\; G_{M N} F^2 \right) \label{sf} \\
& & S_{M N} = T_{M N} - \frac{G_{M N}}{D - 2} \; T
\; \; , \; \; \; \; T = G^{M N} \; T_{M N} \; . \label{smn}
\end{eqnarray} 

Consider the metric $G_{M N}$ of the form given by 
\begin{equation}\label{metric}
d s^2 = g_{\mu \nu} \; d x^\mu d x^\nu 
+ \sigma_{(i)} \; \delta_{i j} \; d y^i d y^j 
+ S \; \omega_{a b} (\theta^c) \; d \theta^a d \theta^b
\end{equation}
where $\mu = 0, 1, \cdots, (d - 1) \;$; $i = 1, 2, \cdots, m
\;$; $a = 1, 2, \cdots, (n + 1) \;$; and $g_{\mu \nu}$,
$\sigma_{(i)}$, and $S$ are functions of $x^\mu$ only and are
independent of $y^i$ and $\theta^a$. Note that the dimensions of
the spacetime is $D = n + m + d + 1$. The Ricci tensor $R_{M N}$
for the above metric can be calculated straightforwardly and its
non zero components can be written as
\begin{eqnarray}
R_{i j} & = & - \; \frac{1}{2} \; G_{i j} \; 
\nabla^2 \; ln \; \sigma_{(i)} \label{i} \\
R_{a b} & = & R_{a b}(\omega) - \frac{1}{2} \; G_{a b} \; 
\nabla^2 \; ln \; S  \label{s} \\
R_{\mu \nu} & = & R_{\mu \nu}(g) - \frac{1}{2} \; 
\nabla^g_\mu \nabla^g_\nu \; \; 
ln \left( \Sigma S^{n + 1} \right) \nonumber \\ 
& & - \frac{1}{4} \sum_{i = 1}^m \left( 
\frac{\sigma_\mu \sigma_\nu}{\sigma^2} \right)_{(i)} 
- \frac{(n + 1)}{4} \; 
\frac{S_\mu \; S_\nu}{S^2} \; . \label{mu} 
\end{eqnarray} 
In the above expressions, 
$(\;)_\mu = \partial_\mu (\;) \;$,  
$\Sigma = \prod_{i = 1}^m \sigma_{(i)} \;$,  
$\nabla^g_\mu$ is the covariant derivative with respect to 
$g_{\mu \nu} \;$,  
$R_{\mu \nu}(g)$ is the Ricci tensor of $g_{\mu \nu} \;$, and 
$R_{a b}(\omega)$ that of $\omega_{a b} \;$. Note that if
$\omega_{a b}$ is the standard metric on a $(n + 1)-$dimensional
sphere of unit radius, as will be the case here, then 
$R_{a b}(\omega) = n \; \omega_{a b}$. Also,  
$\nabla^2 \psi = G^{M N} \nabla_M \nabla_N \psi$ and, for any
function $\psi(x^\mu)$ which depends on $x^\mu$ only, it is
given by
\begin{equation}\label{nabla2}
\nabla^2 \psi = g^{\mu \nu} \;
\left( \nabla^g_\mu \nabla^g_\nu \psi + \frac{1}{2} \; \left(
\frac{\Sigma_\mu}{\Sigma} + \frac{(n + 1) \; S_\mu}{S} \right)
\psi_\nu \right) \; . 
\end{equation}

For our purposes here, $d = 2$ and thus $D = n + m + 3$, 
$x^\mu = (t, R)$ where $R$ is the isotropic radial coordinate,
and $y^i$ are the coordinates of a $m-$dimensional compact
space. Furthermore, we will take $g_{\mu \nu}$ to be diagonal.

Consider electric type $P-$branes with $P \le m$. Then only
$F_{t R 1 2 \cdots P}$ (and related components obtained by
permutations of indices) will be non zero and it depends on
$x^\mu$ only. Equation (\ref{F}) then implies that
\[
F^{t R 1 2 \cdots P} = \frac{Q \; e^{- \lambda \phi}}
{\sqrt{- g \; \Sigma \; S^{n + 1}}} \; \; , \; \; \; 
g = det \; g_{\mu \nu} = g_{t t} \; g_{R R} 
\]
where $Q$ is a constant. It then follows that 
\[
\frac{F^2}{(P + 2) !} = - \; \frac{Q^2 e^{- 2 \lambda \phi}}
{\Sigma_\perp S^{n + 1}} \; \; , \; \; \; 
\Sigma_\perp = \prod_{i = P + 1}^m \sigma_{(i)}
\]
and $F^2_{M N} = \frac{1}{P + 2} \; G_{M N} F^2$ for $MN =
\left( tt, RR, 11, 22, \cdots, PP \right)$ and $= 0$ for other
$(MN)$ components. One then obtains that $S^F_{M N} = G_{M N} \;
S^F_{(M)}$ where
\begin{eqnarray}   
S^F_{(M)} & = & - \; \frac{D - 3 - P}{2 (D - 2)} \; 
\left( \frac{Q^2 e^{- \lambda \phi}}{\Sigma_\perp S^{n + 1}} 
\right) \; \; , \; \; \; M = t, R, 1, 2, \cdots, P \nonumber \\
S^F_{(M)} & = & \frac{P + 1}{2 (D - 2)} \;  \left( 
\frac{Q^2 e^{- \lambda \phi}}{\Sigma_\perp S^{n + 1}} \right) 
\; \; \; \; , \; \; \; \; \; \; M = P + 1, \cdots, m, a \; . 
\end{eqnarray}
Also, equation (\ref{phi}) becomes 
\begin{equation}
\nabla^2 \phi = - \; \frac{\lambda}{2} \; \left( 
\frac{Q^2 e^{- \lambda \phi}}{\Sigma_\perp S^{n + 1}} 
\right) \; .
\end{equation}

We will assume that the energy momentum tensor $T_{M N}$ of the
matter fields is of the perfect fluid form and is given by
\[
T^t_{\; \; t} = - \rho \; , \; \; 
T^i_{\; \; j} = \delta^i_{\; \; j} \; p_{(i)} \; , \; \; 
T^R_{\; \; R} = p \; , \; \; 
T^a_{\; \; b} = \delta^a_{\; \; b} \; p \; .
\]
It then follows that $S_{M N} = G_{M N} \; S_{(M)}$ where 
\begin{eqnarray}
S_{(t)} & = & - \; \frac{1}{D - 2} \; \left( 
(D - 3) \rho + \Pi \right) \label{st} \\ 
S_{(i)} & = & \frac{1}{D - 2} \; \left( 
\rho + (D - 2) p_{(i)} - \Pi  \right) \label{si} \\
S_{(R)} & = & S_{(a)} = \frac{1}{D - 2} \; \left( 
\rho + (D - 2) p - \Pi  \right)  \label{sa} 
\end{eqnarray}
with 
\[
\Pi = \sum_{i = 1}^m p_{(i)} + (n + 2) p \; .
\]

\vspace{4ex}

{\bf 3.}  
In \cite{mcvittie}, McVittie had obtained solutions describing
black holes in an FRW universe. See \cite{nolan, book, sing} for
extensive discussions of these solutions, and \cite{book, sv,
dad, gao} for some further generalisations. Here, we will study
similar brane solutions in a time dependent universe.

We consider here an ansatz similar to that of McVittie solutions
and their higher dimensional generalisations \cite{gao}. The
resulting solutions can be thought of as describing uncharged
black branes in a time dependent universe with perfect fluid
matter. \footnote{McVittie type solutions for Reissner-Nordstrom
black holes in arbitrary dimensions are obtained in \cite{sv,
gao}. Similarly, it should be possible to generalise the ansatz
(\ref{ansatz}) and obtain solutions describing charged black
branes in a time dependent universe with perfect fluid matter.
The ansatz in (\ref{metric}) can indeed include the charged case
but is far too general, and the resulting equations of motion
are difficult to solve directly. Hence an ansatz in between
(\ref{metric}) and (\ref{ansatz}) is needed.  However, a few
straightforward ones we tried did not work and the appropriate
one is not clear to us at present. Hence we will consider only
the uncharged case here.}  In this ansatz, 
$F = 0 \;$, $e^\phi = a^\gamma$, and
\begin{equation}\label{ansatz}
d s^2 = - Z d t^2 + \sum_{i = 1}^m a^{2 \alpha_i} \left( 
d y^i \right)^2 + a^{2 (1 + \beta)} (1 + Y)^{\frac{4}{n}} 
\left( d R^2 + R^2 d \Omega_{n + 1}^2 \right)
\end{equation}
where $a(t)$ is a function of $t$ only,
\begin{equation}
Z = \left( \frac{1 - Y}{1 + Y} \right)^2 
\; \; , \; \; \; Y = \frac{R_0^n}{a^n R^n} \; , 
\end{equation}
$\alpha_i, \beta, \gamma$, and $R_0$ are constants, and 
$d \Omega_{n + 1}$ is the standard line element on an unit 
$(n + 1)-$dimensional sphere. 

The above ansatz is chosen so that for $a = 1$ it describes
static uncharged black brane solutions \cite{solns}.
\footnote{One can include arbitrary powers of $Z$ in various
terms in (\ref{ansatz}) and in $e^\phi$ which, for $a = 1$, will
then describe the multiparameter uncharged solutions given in
\cite{multi}.} Also, for $R_0 = 0$ the ansatz (\ref{ansatz})
describes an FRW universe where $a(t)$ is the scale factor. For
$m = \beta = 0$ and $R_0 \ne 0$, it describes McVittie solution
in $(n + 3)$ dimensions \cite{gao}.

The line element in (\ref{ansatz}) can be written succinctly in
terms of a function $r(R, t)$ defined by
\begin{equation}\label{r}
r(R, t) = a \; R \; (1 + Y)^{\frac{2}{n}} \; .
\end{equation}
It is then easy to verify that 
\begin{equation}\label{Z}
Z = 1 - \frac{r_0^n}{r^n} \; \; , \; \; \; r_0^n = 4 R_0^n 
\end{equation}
and that 
\begin{equation}\label{rtetc}
\frac{r_t}{r} = h \sqrt{Z} \; \; , \; \; \; 
\frac{r_R}{r} = \frac{\sqrt{Z}}{R} \; \; \; ; \; \; \; 
\frac{Z_t}{Z} = h X \; \; , \; \; \; 
\frac{Z_R}{Z} = \frac{X}{R} 
\end{equation}
where we have defined 
\[
(\;)_t = \frac{\partial \; (\;)}{\partial t} \; \; , \; \; \;
(\;)_R = \frac{\partial \; (\;)}{\partial R} \; \; , \; \; \;
h(t) = \frac{a_t}{a} \; \; , \; \; \;
X = \frac{1}{\sqrt{Z}} \; \frac{n \; r_0^n}{r^n} \; .
\]
The ansatz (\ref{ansatz}) now becomes
\begin{equation}\label{final}
d s^2 = - Z d t^2 + \sum_{i = 1}^m a^{2 \alpha_i} \left( 
d y^i \right)^2 + a^{2 \beta} \; r^2 
\left( \frac{d R^2}{R^2} + d \Omega_{n + 1}^2 \right)
\; \; , \; \; \; e^\phi = a^\gamma \;
\end{equation}
where $r(R, t)$ and $Z \left( r(R, t) \right)$ are given in
equations (\ref{r}) and (\ref{Z}).

One can now obtain the equations of motion. Defining
\[
u = \sum_{i = 1}^{m} \alpha_i \; \; , \; \; \; \; 
\tilde{u} = u + (n + 2) \left( \beta + \sqrt{Z} \right) \; ,
\]
the equations of motion (\ref{G}) and (\ref{phi}) can be
written, after a long but straightforward calculation, as 
\begin{eqnarray}
h X \left( u + (n + 1) \beta \right) & = & 0  \nonumber \\
\gamma \left( h_t + ( \tilde{u} - \frac{X}{2} ) \; h^2 
\right) & = & 0 \nonumber \\
\frac{\alpha_i}{Z} \left( h_t + ( \tilde{u} - \frac{X}{2} )
\; h^2 \right) & = & \kappa^2 S_{(i)} \nonumber  \\
\frac{\beta}{Z} \left( h_t + ( \tilde{u} - \frac{X}{2} ) 
\; h^2 \right) + \frac{1}{\sqrt{Z}} \; \left( h_t 
+ \tilde{u} \; h^2 \right) & = & \kappa^2 S_{(a)} \nonumber \\
\frac{h^2}{Z} \; \left( \sum_{i = 1}^m \alpha_i^2 + (n + 2)
(\beta + \sqrt{Z})^2 + \frac{\gamma^2}{2} - \tilde{u}^2 \right)
& = & - 2 \; \kappa^2 \rho  \label{eom}
\end{eqnarray}
where $S_{(i, a)}$ are given by equations (\ref{si}), (\ref{sa})
and we have used $F = 0$ and equations (\ref{i}) $-$ (\ref{mu}).
\footnote{In (\ref{eom}), the second equation is obtained from
equation (\ref{phi}) and the remaining ones from (\ref{G}): the
first, third, and fourth from $M N = (t R, \; i j, \; a b \; \;
or \; \; R R)$ in (\ref{G}) and the last one is a combination of
these and (\ref{G}) with $M N = t t$.} Equations (\ref{t=0})
then follow from equations (\ref{eom}) by Bianchi identities.

Using equations (\ref{G}) and (\ref{st}) $-$ (\ref{sa}), the
Ricci scalar can be written as
\begin{eqnarray}
G^{M N} R_{M N} & = & \frac{1}{2} \; \partial_M \phi \;
\partial^M \phi \; + \; \kappa^2 \; \left( S_{(t)} 
+ \sum_{i = 1}^m S_{(i)} + (n + 2) S_{(a)} \right) \nonumber \\
& = & \frac{1}{2} \; \partial_M \phi \; \partial^M \phi \; 
+ \; 2 \kappa^2 \; \frac{\rho - \Pi}{D - 2} \label{ricci}
\end{eqnarray}
and can be expressed explicitly in terms of the quantities given
in equation (\ref{final}) by using equations (\ref{eom}) and
noting that
\[
\frac{\rho - \Pi}{D - 2} \; = \; \sum_{i = 1}^m S_{(i)} 
+ (n + 2) S_{(a)} \; - \rho \; . 
\]
It follows from this explicit expression that if $r_0 \ne 0$
then, generically, the Ricci scalar diverges as $Z \to 0$. For
$\alpha_i = \beta = \gamma = 0$ this divergence is absent if
$h_t = 0$, whereas for generic values of $(\alpha_i, \beta,
\gamma)$ it is absent only if $h_t = h = 0$. See \cite{sing}
for a detailed study of this singularity for the 
$D = 4$, $n = 1$ case. 

We now consider various limits where equations (\ref{eom})
reduce to the known ones. 

\vspace{2ex}

\noindent
{\bf (I)}
For $\rho = p_i = p = 0$ and $h = 0$, equivalently 
$a = constant$, equations (\ref{eom}) are all satisfied. This is
just the standard static black brane solution. 

\vspace{2ex}

\noindent
{\bf (II)}
Let $r_0 = 0$. Then $Z = 1$, $X = 0$ and $\tilde{u} = 
u + (n + 2) \left( \beta + 1 \right)$. Equations (\ref{eom})
become
\begin{eqnarray}
\gamma \left( h_t + \tilde{u} h^2 \right) & = & 0 \nonumber \\
\alpha_i \left( h_t + \tilde{u} \; h^2 \right) & = & 
\kappa^2 S_{(i)} \nonumber \\
(\beta + 1) \; \left( h_t + \tilde{u} \; h^2 \right) & = &
\kappa^2 S_{(a)} \nonumber \\
h^2 \; \left( \sum_{i = 1}^m \alpha_i^2 + (n + 2) (\beta + 1)^2
+ \frac{\gamma^2}{2} - \tilde{u}^2 \right) & = & 
- 2 \; \kappa^2 \rho  \; \; . \label{frw}
\end{eqnarray}

\vspace{2ex}

\noindent
{\bf (II a)}
Standard FRW equations are obtained by setting $\gamma = 0$,
$p_i = p$ and, with no loss of generality, $\alpha_i = 1$ and
$\beta = 0$ in equations (\ref{frw}). Then 
\[
\tilde{u} = (D - 1) \; \; , \; \; \; 
\Pi = (D - 1) p \; \; , \; \; \; 
S_{(i)} = S_{(a)} = \frac{\rho - p}{D - 2}
\]
and we get 
\begin{eqnarray}
h_t + (D - 1) \; h^2 & = & \frac{\kappa^2 \; (\rho - p)}{D - 2}
\nonumber \\
(D - 1) (D - 2) h^2 & = & 2 \; \kappa^2 \rho \; . 
\label{stdfrw} 
\end{eqnarray}

\vspace{2ex}

\noindent
{\bf (II b)}
When $\rho = p_i = p = 0$ and $a \propto t$, Kasner type vacuum
solutions can be obtained from equations (\ref{frw}) by setting
$n + 2 = 0$ formally and ignoring the $S_{(a)}-$equation. Then
$m = D - 1$ and equations (\ref{frw}) imply that
\[
\tilde{u} = \sum_{i = 1}^{D - 1} \alpha_i = 1 
\; \; , \; \; \; 
\sum_{i = 1}^{D - 1} \alpha^2_i + \frac{\gamma^2}{2} = 1 \; .  
\]
The isotropic vacuum solutions studied in \cite{bagchi} are a
special case of Kasner type solutions where $\alpha_i = \alpha$.
Then
\begin{eqnarray*}
& & \alpha = \frac{1}{D - 1} \; \; , \; \; \; 
\gamma^2 = \frac{2 (D - 2)}{D - 1} \; . 
\end{eqnarray*}

\vspace{2ex}

\noindent
{\bf (III)}
Let $r_0 \ne 0$ and $h \ne 0$. Equations (\ref{eom}) then imply
that, generically,  
\[
u + (n + 1) \beta = \gamma = 0 \; .
\] 
If $S_{(i)} = 0$ also, equivalently $\alpha_i = 0$, for $i = 1,
2, \cdots, m$, then $u = 0$ and, hence, $\beta = 0$. Then, as
follow from equations (\ref{eom}),
\begin{eqnarray*}
\frac{h_t }{\sqrt{Z}} + \; (n + 2) h^2 & = & 
\kappa^2 S_{(a)} = \frac{\kappa^2 \; (\rho - p)}{n + 1} \\
(n + 1) (n + 2) h^2 & = & 2 \kappa^2 \rho 
\end{eqnarray*}
which imply that $\rho$ is homogeneous and is a function of $t$
only. Note that if $m = 0$, {\em i.e.} if there are no compact
directions, then in equations (\ref{eom}) $\alpha_{(i)} = p_i =
0$ formally and $S_{(i)}-$equations are to be ignored. Then
$\beta = 0$ and equations (\ref{eom}) reduce to the above ones
with $D = n + 3$. These are just the higher dimensional,
spatially flat, McVittie solutions given in \cite{gao}.

\vspace{4ex}

{\bf 4.}  
When $a = 1$ the function $Z$ in (\ref{Z}) is time independent
and the ansatz (\ref{ansatz}), equivalently (\ref{final}),
describes uncharged branes with ADM mass $M \propto r_0^n$
\cite{multi}. When $a(t)$ is time dependent, the function $Z$ is
also time dependent and then the ansatz (\ref{ansatz}) can
naturally be thought of as describing uncharged branes with time
dependent mass, the time dependence being dictated by $a(t)$.

In string/M theory context the brane solutions (\ref{ansatz})
with $a = 1$ are often interpreted as coincident stacks of equal
number, $N$, of branes and antibranes with $N \gg 1$ \cite{d}.
\footnote{ More general static multiparameter solutions and
their interpretation in terms of tachyon condensation are
studied in \cite{multi}.} Such a system has a tachyon mode
indicating the instability due to brane antibrane pair
annihilation. As these pairs decay, the tachyon field rolls down
from the maximum of its potential and condenses at the potential
minimum. At the end of this process the brane antibrane pairs
would have annihilated totally and disappeared. \footnote{In the
charged case the numbers $N$ and $\bar{N}$ of branes and
antibranes are not equal and so the decay process will end in
extremal branes with charge $\propto (N - \bar{N})$.}

In recent years, enormous progress has been made in
understanding the tachyon condensation process in the case of a
single brane antibrane pair. See the recent comprehensive review
\cite{ashoke} and references therein. However, not much is known
in the case of stacks of brane antibrane pairs with $N \gg 1$.
Nevertheless, it is reasonable to assume that the dynamics is
qualitatively similar to that of a single brane antibrane pair;
namely, that all the brane antibrane pairs will annihilate
totally at the end.

For $N$ sufficiently large, the above dynamic process must have
a gravity description. During this process $N$, and the mass
$M$, will decrease in time. Such a description must be valid
atleast in the initial stages when $N(t)$ remains sufficiently
large. Then it is natural to hope that an ansatz similar to
(\ref{ansatz}) may describe such a process.

That the ansatz (\ref{ansatz}) can describe time dependent mass
when $a(t)$ is time dependent can be shown by using the
quasilocal Hawking mass \cite{sing, hawking}. Here, for the
purpose of illustration, let us instead imagine that the system
remains static for $t \le 0$ with $a(0) = 1$, that the time
dependence is switched on at time $t = 0_+$, and off at $t =
t_{0_-} > 0$ with $a(t_0) = a_0$, and that the system remains
static again for $t \ge t_0$. \footnote{In the case of tachyon
condensation, switching on the system is analogous to nudging
the tachyon field away from its potential maximum towards the
minimum; switching off, while its meaning is not entirely clear,
is perhaps analogous to the condensation at the potential
minimum if it happens in a finite time.} Let us also assume that
$a_0 \ne 1$ and that the parameters in the ansatz (\ref{ansatz})
obey $u + (n + 1) \beta = 0$ for $t > t_0$, assumptions which
seem physically reasonable.

We then have $(a, h, h_t) = (1, 0, 0)$ for $t < 0$ and $= (a_0,
0, 0)$ for $t > t_0$. The initial and final masses, denoted as
$M$ and $M_0$ respectively, can then be calculated \footnote{The
calculation of $M_0$ is straightforward once one defines
$R_{fin} = a_0^{1 + \beta} \; R$ in (\ref{ansatz}).} in the
standard way \cite{multi} without requiring the concept of
quasilocal mass \cite{sing, hawking}. They are given by
\begin{equation}\label{ms}
M = \frac{(n + 1) \; \omega_{n + 1} \; V_m \; r_0^n} 
{2 \kappa^2} \; \; , \; \; \; 
M_0 = a_0^{u + n \beta} \; M = \frac{M}{a_0^\beta}
\end{equation}
where $\omega_{n + 1}$ is the area of an unit $(n +
1)-$dimensional sphere, $V_m$ is the initial volume of the
$m-$dimensional compact space, and the relation $u + (n + 1)
\beta = 0$ is used. So, the final mass $M_0$ is different from
the initial mass $M$ if $a_0^\beta \ne 1$. Thus if $a_0^\beta >
1$ then $M_0 < M$ indicating that the final mass decreases. Note
that if $u = 0$, which is necessarily the case if $m = 0$, then
$\beta = 0$ and $M_0 = M$. This is the case for McVittie
solutions since $m = u = 0$ there \cite{mcvittie, dad, gao}.

Bekenstein-Hawking entropies $S_{BH}$ and $S_{0BH}$, for $t \le
0$ and $t \ge t_0$ respectively, can also be calculated and are
given by
\begin{equation}\label{sbh} 
S_{BH} = \frac{2 \pi \; \omega_{n + 1} \; V_m \; r_0^{n + 1}}
{\kappa^2} \; \; , \; \; \; 
S_{0BH} = a_0^{u + (n + 1) \beta} \; S_{BH} = S_{BH} \; .
\end{equation} 
Thus the final entropy $S_{0BH}$ and initial entropy $S_{BH}$
are equal here which is a consequence of the relation $u + (n +
1) \beta = 0$ which, in turn, is a consequence of the absence of
the source term $S_{t R}$ as can be seen from (\ref{eom}). This
indicates that a non zero source term $S_{t R}$ is likely to be
necessary to model a system such as decaying brane antibrane
pairs where entropy decrease is expected on physical grounds.
However, a non zero $S_{t R}$ alone may not be sufficient, see
the discussion below in section {\bf 5}.

\vspace{4ex}

{\bf 5.}  
Clearly a more general ansatz is needed to model brane antibrane
decay process, equivalently tachyon condensation. We now discuss
the necessary generalisations which, hopefully, are also
sufficient.

In the decay of a single brane antibrane pair ($N = 1$), the
tachyon field resides in the brane world volume. Also, in the
weak coupling limit where the string coupling constant $g_s \to
0$, the total energy on the brane remains constant as the
tachyon rolls down from the potential maximum to the minimum.
This is because the emission of closed string radiation is
suppressed in the limit $g_s \to 0$. However, for non zero
$g_s$, it is expected that closed strings are radiated into the
transverse non compact space, thereby decreasing the total
energy on the brane. For more details see \cite{ashoke} and
references therein.

Assuming that similar qualitative features persist also for the
case when $N \gg 1$ and $g_s$ is non vanishing then suggests the
following generalisations:

\vspace{2ex}

\noindent  
$\bullet$ 
A delta function source $S_\parallel \; \delta (R_{trans})$ must
be included with $S_\parallel$ given by the energy momentum
tensor of a world volume scalar field $T$ rolling down its
potential $V(T)$. \footnote{ The source term and the energy
momentum tensor are related as in equation (\ref{smn}).} Note
that, even for the $N = 1$, $g_s \to 0$ case, $V(T)$ is not
known from first principles. For the $N \gg 1$ and non vanishing
$g_s$ case required here, one has to first find $V(T)$ from
first principles or, as in the $N = 1$ case, find an empirical
form with correct properties.

Instead of a delta function source one may include, more
generally, a smeared source localised close to the brane world
volume and assume different metrics, {\em e.g.} the metric in
(\ref{metric}) with $(g_{\mu \nu}, \sigma_{(i)}, S)_{in}$ and
$(g_{\mu \nu}, \sigma_{(i)}, S)_{out}$, and different source
terms in this region and in the region outside it. Appropriate
junction conditions are then to be imposed to ensure continuity
across the boundary. See \cite{nolan} for an example.

\vspace{2ex}

\noindent
$\bullet$
Source terms representing outgoing radiation in the transverse
space must be included. Such terms will have non zero $S_{t R}$.
It may then be possible to have $S_{0BH} < S_{BH}$, see equation
(\ref{sbh}) and related comments. 

\vspace{2ex}

\noindent
$\bullet$ 
Area increase theorems of General Relativity \cite{he} forbid
the decrease of horizon area and hence of Bekenstein-Hawking
entropy $S_{BH}$. Now, if the brane decay dynamics is to result
in the decrease of horizon area, and thus of $S_{BH}$, 
as seems physicaly reasonable then these theorems must be
invalidated, presumably by violating some energy conditions
assumed in their proofs.

It is not clear if the outgoing radiation terms, mentioned
above, will suffice. So, in general, one may include empirically
a perfect fluid source with an appropriate equation of state so
that some combination of outgoing radiation and this perfect
fluid may invalidate the area increase theorems and result in
the decrease of $S_{BH}$. In this context note that black hole
mass and, consequently, entropy is known to decrease due to the
accretion of phantom energy (a perfect fluid with
$\frac{p}{\rho} < - 1$) which violates energy conditions, see
\cite{babichev} and references therein. It will be quite
interesting if similar energy source(s) were to originate in
string theory in the context of brane decay. 

\vspace{2ex}

In conclusion, we note that incorporating the above
generalisations in the ansatz for the metric and the source
terms $S_{M N}$ is comparatively easy. Indeed the ansatz given
in equation (\ref{metric}) for the metric is quite general and
suffices also when the above generalisations are incorporated.
However, it is a tall order to solve the resulting equations of
motion, extract the salient features, and interpret them in
terms of the dynamics of brane decay, equivalently tachyon
condensation, in string theory. Nevertheless such a project
seems worthwhile in view of the insights it is likely to offer.



\end{document}